# Piezoelectric SPDT NEMS Switch for Complementary Logic

Xilin Liu, Haodi Ouyang, Yidong Wang
*Abstract*—**Off-state current leakage and switching delay has become the main challenge for continued complementary metal-oxide-semiconductor (CMOS) technology scaling. Previous work proposes a "see-saw" relay structure to mimic the operation of CMOS. This paper presents a novel single-pole double throw (SPDT) switch structure based on AlN piezoelectric cantilever beam to improve the former "see-saw" relay structure. Geometry parameters are given and key switch parameters such as actuation voltage, switching time and contact force have been calculated and compared with previous "see-saw" relay structure. Analysis and design process is shown and micro-fabrication process is described as well.**

## I. Introduction

Over the past 40 years, large progress in integrated circuits (IC) technology has been made. The key to the continual improvements has been the steady miniaturization of metal oxide semiconductor field-effect transistor (MOSFET), the basic building block used predominantly in IC chips today. However, power density has grown to be the dominant challenge for CMOS scaling because of the off-state leakage current between the drain and source increases exponentially with decreasing voltage [1].

In contrast, mechanical switch is ideal in that it has essentially zero off-state current and hence eliminates the trade-off between decreasing active energy and increasing leakage energy [2]. It also has abrupt switching behavior, which allows $V_{ON}$ to be decreased while maintaining relatively high On-state current [1]. An example of this is the "see-saw" structure which is suspended by two torsion beams allowing the ends of the gate to move up and down in a perfectly complementary fashion [3].

Many electrostatic actuation-type RF MEMS switches have been developed for their easy fabrication process and good dynamic characteristics. But their operation voltages were several tens volts. If the thickness of a cantilever is decreased to lower the spring constant and actuation voltage, the restoring force of the cantilever becomes low too. This hinders the switch from restoring and makes the cantilever stick to a substrate at a contact point [3].

Therefore it is necessary to develop piezoelectric switches. The main benefit of the piezoelectric MEMS switch is that low voltage operation is possible, so the reliability can be improved by eliminating the sticking problem at contact points by using of an active "OFF" operation.

In this paper, we have designed a novel piezoelectric SPDT mechanical switch. It composes of a cantilever beam actuated by piezoelectric force which can reasonably reduce the actuate voltage.

## II. Design and Principle of Operation

The basic structure and operation of our piezoelectric SPDT NEMS switch is shown in fig. 1. The main beam is formed by two layers of piezoelectric film undergoing opposite deformations. The operation is shown is in fig. 2. When a positive voltage is applied on the middle Platinum electrode and other two Platinum electrodes grounded, the top AlN layer will contract, generating a horizontal force, and the bottom AlN layer will expend, generating an opposite force. The bending moment has an equivalent force up, thus the beam will bend up; On the other hand, when a negative voltage is applied, the beam will bend down. By carefully design the biasing and threshold voltage, the beam will bend up on "1" logic voltage, and bend down on "0" logic voltage.

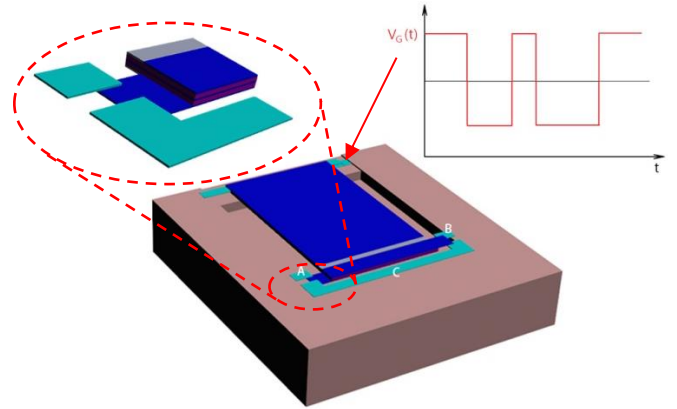

Fig. 1. 3-D schematic view of the piezoelectric SPDT NEMS switch. The picture illustrates the position of the contact and the electrodes. Zoom part is the contact on the bottom left. $V_G(t)$ shows an example of input signal in voltage after biasing.

There are two contacts working as the "double throw", one on the top to the right, and the other one on the bottom to the left. The contacts are made of Platinum and isolated from the aforementioned Platinum electrode layers. When the beam bends up, the electrodes A and C get short; and when the beam bends down, the electrodes B and C get short. In application, electrodes A and B are set to be constant voltage or as input port; electrode C works as the output port. An "And" logic realized in this structure is shown in fig. 3. It is apparent that this structure can realized all the logic that a previous "see-saw" structure achieved in mimicking the operation of logic circuit [2].

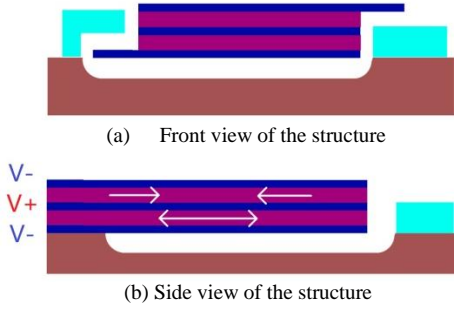

(a) Front view of the structure

(b) Side view of the structure

Fig. 2. Front and side schematic view of the piezoelectric SPDT NEMS switch showing the principle of operation.

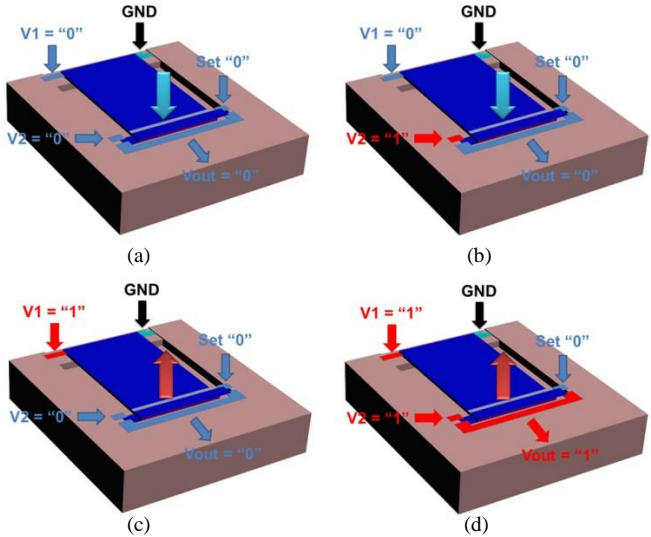

Fig. 3. "AND" logic realized by this piezoelectric SPDT NEMS switch shown in 3-D schematic view.

### III. ANALYSIS AND DISCUSSION

As mentioned before, the conventional mechanical actuation mechanisms, like capacitive that have been used to drive nano-scale devices have an obvious advantage that it cannot scale well in the sub-micron dimensions [4].

Besides, piezoelectric devices can take advantage of a reverse potential to generate higher forces to open the switch. What's more, it requires lower power for operation [5].

To optimize our parameters we need to make trade-offs in several aspects. According to the principle of operation, the piezoelectric force (horizontal) generated in both active layers by a certain applied voltage is:

$$F_p = d_{31}E_{AlN}WV_{app} \quad (1)$$

thus the total vertical force generated by the applied voltage can be calculated by equivalent moment:

$$F_{tot} = \frac{d_{31}E_Y(WT)}{2L}V_{app} \quad (2)$$

The contact force equals to the vertical force minus the mechanical force caused by closing the gap. The stiffness of the cantilever beam can be estimated according to the geometry parameters [6]:

$$K_{me} = \frac{E_{eq}W}{4}(\frac{T}{L})^3 \quad (3)$$

Thus the contact force is given by:

$$F_{contact} = \frac{d_{31}E_Y(WT)}{2L}V_{app} - \frac{E_{eq}W}{4}(\frac{T}{L})^3 gap \quad (4)$$

According to Maxwell's (or Sharvin's) model [7], the contact resistance R is inverse proportional to the radius of curvature "r" in the contact area, while "r" is proportional to force ($r \propto F$ $r \propto \sqrt{F}$). So it is desirable to increase the contact force thus having a low contact resistance.

The relation of Beam length and contact force is shown in fig. 4 and fig. 5. For a given thickness, the contact force increases to the maximum value as the beam length increases, and then decreases; a smaller thickness will lead to a shorter length when the maximum force is reached. So a smaller thickness is desirable. However, according to the state-of-the-art micro fabrication process [4], a thickness of 350nm of the whole beam seems to be the limit (3 layers of 50nm Platinum, 2 layers of 100 nm AlN).

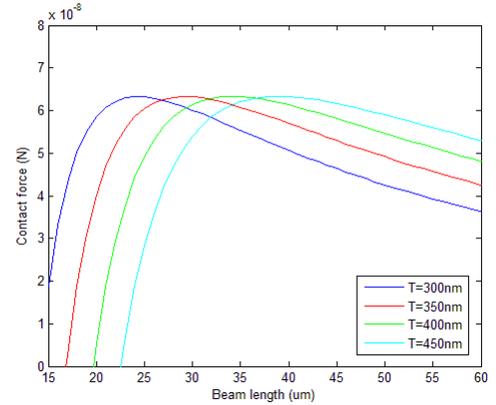

Fig. 4. The relation of beam length and contact force given gap size 40 nm, width 20um, applied voltage 2.5V

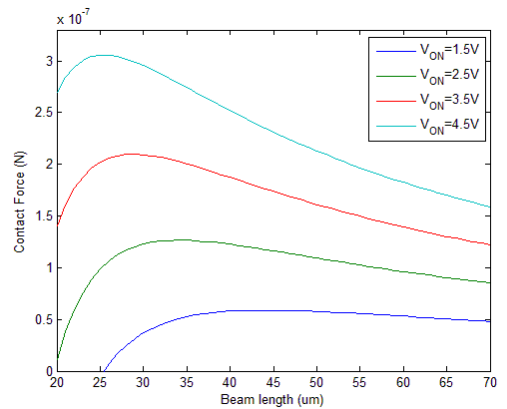

Fig. 5. The relation of beam length and contact force given gap size 40 nm, width 20um, thickness 350nm

Assume a thickness of 350nm, the relation between beam length, applied voltage and contact force is shown in fig. 5. It is apparent that a higher applied voltage will lead to a higher maximum contact force with smaller beam length. However, lower voltage is often required by logic circuit.

The natural and damped resonant frequencies of a cantilever beam are given by equation (5) and (6):

$$\omega_0 = \frac{1.875^2}{\sqrt{12}} \frac{T}{L^2} \sqrt{\frac{E_{eq}}{\rho}} \quad (5)$$

$$\omega_d = \omega_0 \sqrt{1-(\frac{1}{2Q})^2} \quad (6)$$

The displacement at the end of the beam is decided by the characteristics of the beam and the applied force.

$$x = \frac{F_{tot}}{m\omega_0^2}[1-\frac{1}{\omega_d}\exp(-\frac{\omega_0 t}{2Q})(\frac{\omega_0 t}{2Q}\sin(\omega_d t)+\omega_d \cos(\omega_d t))] \quad (7)$$

Taking into account just the first order term we have:

$$t_s = \frac{2Q}{\omega}\sqrt{\frac{k}{F_{tot}}gap} \quad (8)$$

Fig. 6 and 7 shows the relation between switching time, applied voltage and displacement. In order to decrease the switching time, it is desirable to have smaller gap size and larger applied voltage. According to the recent micro fabrication technology, we choose 40nm as the gap size in our design.

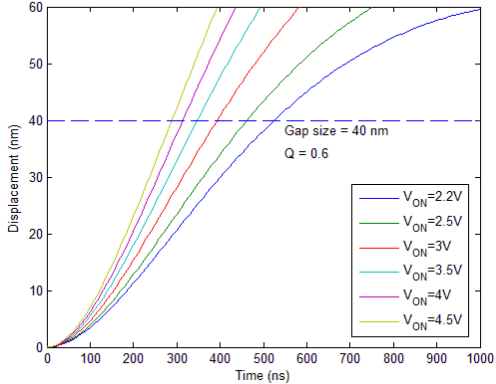

Fig. 6. The relation between switching time and displacement according to equation (7) given gap size 40 nm, length 30um, width 20um, thickness 350nm

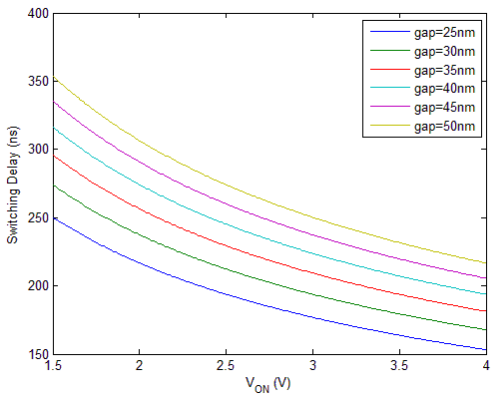

Fig. 7. The relation between switching time and applied voltage according to equation (8) given beam length 30um, width 20um, thickness 350nm

The applied force should be finally decided by the contact resistance. We choose an actuating voltage of 5 V ($\pm$2.5V on both sides). According to Maxwell's spreading resistance equation, the radius of curvature in our design is about 135.53nm, larger than the electron mean free path, thus:

$$R = \frac{\rho}{2r} \quad (9)$$

The estimated contact resistance is around 0.51 Ω in our design, which satisfies the general require of NEMS switch.

In sum, after considering all these models and tradeoff above, we choose the parameters as shown in table I. Compared with latest "see-saw" logic gate [8], which cover an area of almost 4000 μm² our design limit the cover area to about 600 μm². And our actuating voltage is lower than that of "see-saw" structure [7], which requires 7V (body to source voltage).

TABLE I
DESIGN PARAMETERS OF THE SPDT NEMS SWITCH

| Symbol | Parameter | Quantity | Unit |
|---|---|---|---|
| $L$ | Length | 30 | μm |
| $W$ | Width | 20 | μm |
| $T$ | Thickness | 350 | nm |
| $gap$ | Initial Gap | 40 | nm |
| $K_{me}$ | Mechanical Stiffness | 3.09 | N/m |
| $F_{tot}$ | Total Generated Force | 0.215 | μN |
| $F_{contact}$ | Contact Force | 0.123 | μN |
| $t_s$ | Switching Time | 212 | ns |
| $Q$ | Quality Factor | 0.6 | — |
| $R$ | Contact Force | 0.51 | Ω |

## IV. FABRICATION

An eight-mask process is used for the fabrication of the switch, as shown in fig. 8. And all the masks used are shown in fig. 9.

First, a layer of Platinum (50 nm) is deposited on the silicon and patterned by lift-off. Then the sacrificial layer (40 nm) is deposited on the left electrode of the beam to produce the air gap. (Just on the contact electrode area, not the whole electrode). Then the drain and the source of the switch are electroplated on the silicon layer. Then Platinum seed layer and a thick Photo resist are used to mold those Platinum electrodes (50 nm). And on the first layer of the AlN (100 nm), a second AlN (100 nm) is deposited on top of the second layer of the Platinum (50 nm). $SiO_2$ can be used as hard mask and a $Cl_2$ based dry etch process to pattern the AlN layers. Then another sacrificial layer (40 nm) is used at the right of the beam electrode. After that, the third layer of the Platinum (50 nm) can be deposited on the top of the second AlN (100 nm). Then the entire beam structure layers are fabricated. Finally, the sacrificial layers along with the silicon are dry released by $XeF_2$. Dry release is used to eliminate stiction problems that would have otherwise been encountered with wet release techniques [9]. The released structure can be seen in fig. 2.

In general, compared to the MEMS switch, the scaling of the Al film and the devices need a more accurate control of the process parameters. However, on the other hand, the use of thin film simplifies the dry etch step ($Cl_2/BCl_3$ based dry etch of AlN), which does not require the use of an oxide hard mask, and can be performed with hard-baked photoresist [10].

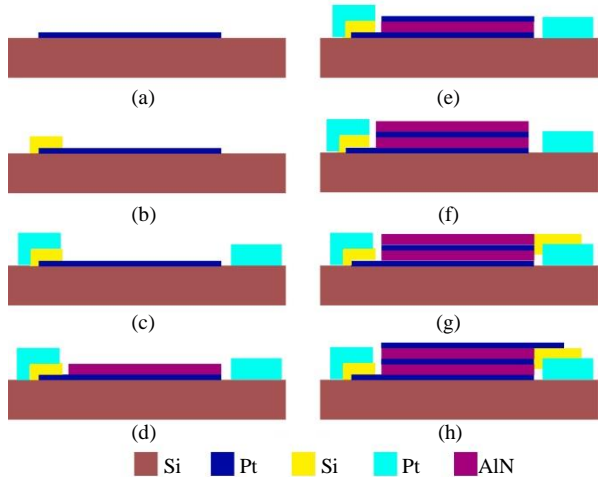

Fig. 8. Process flow of the SPDT NEMS switch shown in front view

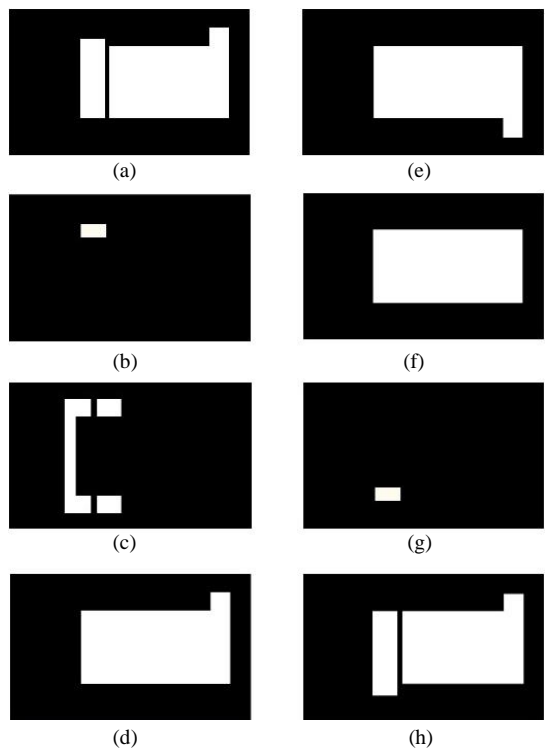

Fig. 9. All the masks used in the microfabrication process

## V. CONCLUSION

In this paper, we put forward a novel piezoelectric SPDT NEMS switch design to implement complementary logic circuit with extremely small size, low operating voltage, and low switching delay. With further design and process improvements, the design shows a promising use for energy-efficient information processing in the future.

ACKNOWLEDGMENT

The authors would like to thank Prof. Piazza for helpful advices. And thank for Siddhartha Ghosh for his kindly guidance in our learning process.